\documentclass[amssymb,prd,aps,showpacs,nofootinbib,preprint]{revtex4-1}
\usepackage{latexsym}
\usepackage{graphicx,color}
\usepackage{graphicx}
\usepackage{amsmath,amssymb}
\usepackage{amsmath}
\usepackage[T1]{fontenc}
\usepackage[hang,small,up]{caption}

\makeindex
\begin{document}
\title{Pattern transitions in a nonlocal logistic map for populations}
\author{Fernando V.  Barbosa$^{1,2,3}$}\author{Andr\'{e} A. L. Penna$^{2,3}$}\author{Rogelma M.S. Ferreira$^{2,3,4}$}\author{Keila L. V. Novais$^{2,3}$}\author{Jefferson A. R. da Cunha$^5$} \author{Fernando A. Oliveira$^{2,3}$}\email{faooliveira@gmail.com}
\affiliation{Instituto Federal de Bras\'{i}lia, Campus S\~{a}o Sebasti\~{a}o$^{1}$\\
Instituto de F\'{i}sica - Universidade de Bras\'{i}lia, Brazil$^{2}$\\
International Center for Condensed Matter Physics\\CP 04455.
70919-970 Bras\'{i}lia DF, Brazil,$^{3}$\\
Centro de Ci\^{e}ncias Exatas e Tecnol\'{o}gicas, Universidade Federal do Rec\^{o}ncavo da Bahia, 44380-000 Cruz das Almas, Bahia, Brazil$^{4}$\\
Instituto de F\'{i}sica, Universidade Federal  de Goi\'{a}s, CP 131 CEP 74001-970, Goi\^{a}nia, Brazil$^5$ }


\begin{abstract}
\noindent

In this work, we study the pattern solutions of doubly nonlocal logistic map that include spatial kernels in both growth and competition terms. We show that this map includes as a particular case the nonlocal Fisher-Kolmogorov equation, and we demonstrate the existence of three kinds of stationary nonlinear solutions: one uniform,  one cosine type that we refer to as wavelike solution, and another in the form of Gaussian. We also obtain analytical expressions that describe the nonlinear pattern behavior in the system,  and we establish the stability criterion. We define thermodynamics grandeurs  such as entropy and the order parameter. Based on this, the pattern-no-pattern and pattern-pattern transitions are properly analyzed. We show that these pattern solutions may be related to the recently observed peak adding phenomenon in nonlinear optics.

\end{abstract}
\date{\today }
  \maketitle

\section{Introduction}

Patterns are spatially organized structures in  nature that arise from the aggregation of elements in many different processes \cite{Fehsenfeld,Kawczynski,Vanag,Barthelemy,Fuentes03,Anna}. Experiments and theory have shown that patterns are related to spatiotemporal nonlinear dynamics equations where specific structures can undergo nucleation \cite{Oliveira96,Gonzalez99}, aggregation, clustering, complex self-cooperative mechanisms, morphology changes, and complex biological variations \cite{Cross,Kenah,Newman,Pamplona,Colombo,Zhabotinsky,Legawiec,Fuentes4,Showalter,Jefferson,Jefferson11}, to mention just a few. The inclusion of nonlocal interactions to study pattern formation has been of great importance  in the explanation  of interactions between neurons, granular systems, and population models with complex dynamics, such as predator-prey interactions \cite{Wakano,Murray,Hallatschek09,Malchow,Kosinski,Biham,Szabo,Keshet}. Nonlocality in space has been implemented through integro-differential reaction-diffusion equations in many physical systems. For example, inter-cellular nutrients and their interactions can be described as nonlocal reaction diffusion effects in which patterns arise when the activator and inhibitor are not present  uniformly in space. Moreover, nonlocal reaction diffusion models have been widely studied in the context of the formation of travelling waves \cite{Hildebrand},  the dynamics of fronts and defects \cite{Nicola},  chemical interactions and Hopf bifurcation \cite{Rovinsky}.

In this study, we propose a nonlocal logistic map that generalizes the Fisher-Kolmogorov equation for populations, whose dynamics is given by 
\begin{eqnarray}
\label{interaction_competition}
\frac{\partial u(x,t)}{\partial t}=a\int_{\Omega}g_{\alpha}(x-x')u(x',t)dx'
-bu(x,t)\int_{\Omega}h_{\beta}(x-x')u(x',t)dx'\,\,,
\end{eqnarray}
where $u(x,t)$ is the population density while the constants $a$ and $b$ refer to the growth and competition parameters, respectively. The kernels $g_{\alpha}(x-x')$ and $h_{\beta}(x-x')$ are the nonlocal correlations for the growth and competition terms, respectively, which weight the interaction among the individuals of the system. The key feature here is that the nonlocal correlations are controlled by parameters $\alpha$ and $\beta$ that refer to the growth and interaction lengths respectively. Note that this allows us to express the effects of nonlocality of growth and interaction using only the lengths $\alpha$ and $\beta$. Note also that Eq.~(\ref{interaction_competition}) incorporates the classical logistic equation $\partial u(x,t)/\partial t = au(x,t)-bu^{2}(x,t)$ when we take the kernels  $h_{\beta}(x-x')=g_{\alpha}(x-x')=\delta(x-x')$ as a delta function. Moreover, if we fix the second integral, another important feature is that the nonlocal growth integral in Eq.~(\ref{interaction_competition}), i.e.,
$I_{\alpha}(x)=\int g_{\alpha}(x-x')u(x',t)dx = \int g(y)u(x+y)dy$, concentrates, in a simple manner, typical growth, diffusion, and higher order derivatives  related to the dispersive terms. This is easily obtained by expanding $u(x+y)$ in Taylor series around $y=0$, from which we obtain  the Fisher-Kolmogorov equation with an kernel for the competition, a growth,  and a diffusive term of the form $D\nabla^2u(x,t) $.  As well containing the higher order derivatives, i.e.,  dispersive terms, which are not included in the one kernel Fisher-Kolmogorov equation, this is an extra gain of the formulation. Moreover, all the properties of these terms are included in the model by varying only the lengths $\alpha$. In this approach the diffusive constant is
\begin{equation}
\label{Dif}
D=\frac{a}{2}\int_{\Omega}g_\alpha(y)y^2dy,
\end{equation}
i.e., it is directly associated with the growth rate and the dispersion of the individuals. Higher order moments can be computed in the same way. Note here that these constants are fixed when  $g_{\alpha}(x)$ is defined.

There many situations in nonlinear dynamics where the evolution to a stationary state is more important than the dynamics itself: for example, the synchronization of nonlinear oscillators is an important cooperative phenomenon that is widely applied in different disciplines, ranging from physics to the social sciences \cite{Strogatz03,Longa96,Bocaletti02,Ciesla01,Morgado07,Acebron05,Hong07,Bonilla93,Sonnenschein13,Park96,Reimann99,Yu13}. The advent of the phase reduction method has  facilitated the deduction of  simpler equations for the study of limit cycle oscillators \cite{Kuramoto84,Nakao07,Zhou02}, enabling a breakthrough in understanding the application of synchronization phenomenon. In the synchronization process the final stationary state is more important, and for that it is possible to define an order parameter. Recently \cite{Pinto16} we have  elaborated a thermodynamics analysis of the synchronization process. Our aim is to develop a similar approach for population dynamics.

The dynamics that can be obtained from the above equation is very rich and our intention is to explore it in future works. However, the study of stationary pattern solutions of Eq.~(\ref{interaction_competition}) is a first step to a more complex dynamical analysis, and it allows us to have a first idea of the pattern formation that can be obtained from a population equation and a stronger analogy with the thermodynamics of phase transitions.
It is possible to solve it analytically and numerically  and express the solution in terms of the parameters $\alpha$ and $\beta$.
We start to show that there are three types of possible solutions; one is the uniform solution $u(x)=constant $, which we do not associate with a pattern, and the other two are pattern solutions in the form of wavelike and Gaussian. We show that these solutions encompass  the properties of nonlocality and nonlinearity of the system in which analytical expressions are presented. These solutions allow us to form an overview of the pattern formation process, including a complete analysis of the pattern focused on the number of peaks and the pattern transitions when we vary $\alpha$ and $\beta$. Based on this, we analyze  the relationship between patterns and peaks, and how the transitions occur.

\section{Pattern solutions}

The nonlocal Eq.~(\ref{interaction_competition}) provides a general model to describe complex population systems that permit us to study its evolution in  space and time through a rapid implementation of the simulation process. Pattern solutions can be obtained
with respect to the spatial inhomogeneous distribution of the system related directly to the length interactions $\alpha$ and $\beta$ among the individuals.

In order to obtain solutions for Eq.~(\ref{interaction_competition}), we need  to define the two kernels; $h_{\alpha}$ and  $g_{\alpha}$. In stochastic processes temporal kernels can be generated via the fluctuation dissipation theorem \cite{Lapas08,Ferreira11}. In our case, we have no equivalent theorem, an thus we have to perform an ad hoc assumption. In order to make comparisons between $\alpha$ and $\beta$, we take
 $h_{\alpha}(y) = g_{\alpha}(y)$ with different correlation lengths $\alpha$ and $\beta$ and then use the step function
\begin{equation}
g_{\alpha}(x)=\begin{cases} \frac{1}{2\alpha}, & \mbox{if } |x|<\alpha \\ 0, & \mbox{otherwise}.
\end{cases}
\end{equation}
This function is easy to operate and give us a first approximation of the correlation behavior. For example if we use it in Eq. (\ref{Dif}) we obtain $D=\frac{1}{6}a\alpha^2$. The diffusion is  proportional to the growth rate, which forces migration, and to the square of the correlation length $\alpha$. A large $\alpha$ means a loose bound between the members of a species that will  desegregate it. In this way, we can associate  $\alpha$ with disorder, and if we compare it with the particles diffusion constant $D \propto k_BT $ we have $ \alpha \propto T^{1/2}$, a good analogy for small values of $\alpha$.

 To solve Eq. (\ref{interaction_competition}) we use the periodic boundary conditions $u(x+L,t)=u(x,t)$. Lengths  $x$, $\alpha$, and $\beta$ are in units of $L$, and the density $u(x,t)$ is in units of $L^{-1}$. Note that the stability of the solution is controlled by the parameter
\begin{equation}
\label{stability}
\eta= \int_0^L[u(x,t+\Delta t)-u(x,t)]^2dx\,.
\end{equation}
By assuming the following numerical conditions: $a=b=1$, $\Delta x=10^{-5}$, and
$\Delta t=10^{-2}$, we solve Eq.~(\ref{interaction_competition}). A convergence has been observed after $n=60000$ time steps with a desired precision $\eta < 10^{-16}$. In this situation we say that Eq.~(\ref{interaction_competition}) reached the steady state solution $U(x)$, i.e.  $U(x)=u(x,t\rightarrow \infty)$.

In Fig. 1 we show the behavior  of the steady state population density $U(x)$ as a function of $x$. The values are obtained by  numerically solving  Eq.~(\ref{interaction_competition}) in the domain $0<x<L$. Here we have fixed, for all curves, the parameter $\alpha=0.03$, while we vary the parameter $\beta$ as (a) $\beta=0.117$; (b) $\beta=0.118$; (c) $\beta=0.1195$. We show here only distributions that originate patterns, i.e., the  non-uniform distributions. We should note that, in all situations, as long as  $u(x,0) >  0$ the final stationary state is independent of the initial distribution $u(x,0)$. In curves (a) and (c) we see the presence of $6$ and $5$ crests, respectively. In particular, for curve (b) we see  a beautiful interference phenomenon that occurs for intermediate values of $\beta$. Altogether, the three curves show the pattern transition from six to five peaks as we increase $\beta$.
The interrupted lines are fitted using the function $U(x)=A[1+Bcos(kx)]$. We see that the fit perfectly match the data, and they are the starting point for our analytical analysis.

\begin{figure}[h!]
\begin{centering}
\rotatebox{-90}{\resizebox{9.0cm}{!}{\includegraphics{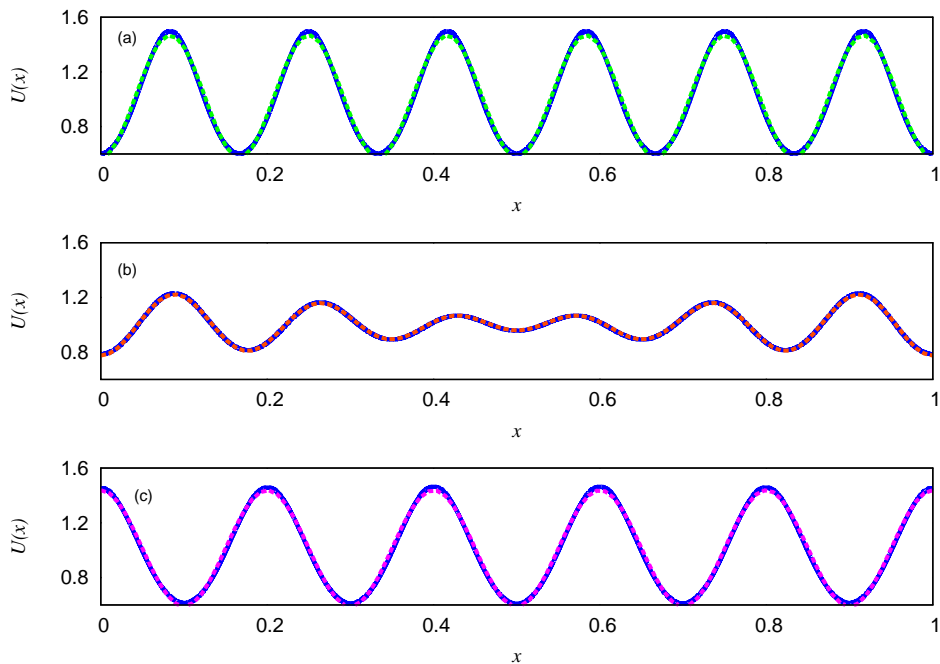}}}
\caption{Population density $U(x)$ as a function of the position $x$. Lengths are given in units of $L$, and $U(x)$ in units of $L^{-1}$, and $\alpha=0.03$ for all curves. Curve (a): $\beta=0.117$; curve (b): $\beta=0.118$; curve (c): $\beta=0.1195$.}
\label{fig.ux}
\end{centering}
\end{figure}

Nonlinear dynamics are of great interest \cite{Murray,Hallatschek09}; nevertheless, in many situations we are interested in verifying if the system evolves for stationary states. We look for pattern solutions in Eq.~(\ref{interaction_competition}) considering the steady state condition $\partial u/\partial t=0$ in the system. It follows that
\begin{equation}
\label{iterative process}
U(x)=\frac{a\int_{\Omega}g_{\alpha}(x-x')U(x')dx'}{b\int_{\Omega}g_{\beta}(x-x') U(x')dx'}\,.
\end{equation}
Note that after convergence, we have dropped the variable $t$ in $u(x,t)$ which means that we have arrived at the steady state $U(x)$. The above equation can be solved iteratively, and we obtain the same result and precision as before. However, the  convergence is faster, and we need now around 200 steps.  Moreover, this provides us a method with to obtain precise analytical pattern solutions. It is easy to see that taking $\alpha=\beta$, Eq.~ (\ref{iterative process}) yields the no pattern solution
\begin{equation}
\label{uni}
U_{NP}(x)=u_0=\frac{a}{b}=1,
\end{equation}
that is the homogeneous or uniform density distribution $u_0$ of the population. Nevertheless, for specific values of  $\alpha$ and $\beta$ the steady state is a non-uniform density distribution that necessarily results in pattern solutions. When we consider non-uniform values of population density in space Eq.~(\ref{interaction_competition}), allows us two types of pattern solutions that we denominate as wavelike and Gaussian. We will examine these solutions in the following sections.

\section{Wavelike patterns}

\subsection{Stability}

Before we present the nonlinear solution we look for the modes stabilities. Consider
\begin{equation}
u(x,t)= u_0+ \varepsilon \exp{(k_m x +\varphi t)}\,,
\label{Landau}
\end{equation}
the second term is a perturbation in the uniform density $u_0$, where $|\varepsilon|\ll 1$, $k_{m}$ is the wavenumber
\begin{equation}
k_{m}=\frac{2\pi m}{L}\,,
\end{equation}
in which $m$ is the crest number and we have used periodic boundary conditions.
Here $\varphi$ is a complex number. If we substitute Eq. (\ref{Landau}) into Eq.~(\ref{interaction_competition}) we obtain
\begin{equation}
\label{lins}
\psi=Re(\varphi)=f_\alpha-f_\beta-1\,,
\end{equation}
where
\begin{equation}
f_{\sigma}=\frac{\sin{k_{m}\sigma}}{k_{m}\sigma}\,,
\end{equation}
 The perturbations are attenuated for $\psi<0$.
Therefore, for any value of $(\alpha,\beta)$, the perturbation will drive  away the system from the uniform distribution if $\psi >0$. This implies that $ f_{\alpha} \approx 1$ and $f_\beta <0$, which naturally requires $\alpha < \beta$. Nevertheless, this solution is limited for small perturbation effects on the nonlocal behavior of Eq.~(\ref{interaction_competition}). Next, we shall propose  nonlinear solutions for the model.

\subsection{Nonlinear solution}

The wave solutions above show that patterns as well as transitions between patterns arise when we change the length $\alpha$ and $\beta$. In light of this, we have denominated these solutions of wavelike patterns. In order to understand this pattern behavior for the wavelike stationary solution we propose a solution of the form
\begin{equation}
\label{cos2}
U(x) = A(m)\left[1+ \sum_{n=1}^{\infty}B_{n}(m)\cos{(nk_m x)}\right],
\end{equation}
where $A(m)$ and $B_{n}(m)$ are functions of $\alpha$ and $\beta$ to be analytically determined. Here $m$ is the number of maxima,  crests, the wave has. We have numerically verified ( see Fig. 1), that for $|B_{1}(m)|< 1$ we get a very good approximation of Eq. (\ref{cos2}) as
\begin{equation}
\label{cos}
U(x) = A(m)\left[1+ B(m)\cos{(k_m x)}\right]\,,
\end{equation}
where we take $B(m)=B_{1}(m)$ for simplicity. Hence, by substituting Eq. (\ref{cos}) into Eq. (\ref{iterative process}), we obtain three possible solutions: one trivial solution with $B(m)=0$ and $A(m)=1$, i.e., which leads to the already well-known homogeneous solution $U(x)=u_0=a/b=1$, with no pattern, and a general solution for $A(m) \neq 1$ and $B(m) \neq 0$, which yields wavelike patterns.
We can determine the parameters $A(m)$ and $B(m)$, which results in
\begin{eqnarray}
A(m)&=&1+\zeta \left(1-\frac{1}{\sqrt{1-z^2}}\right)\\
\nonumber\\
B(m)&=&\pm\frac{1}{2(1+\zeta)f_\beta}\left[4(1+\zeta)^2-\left(\sqrt{1+4\zeta(1+\zeta)(1+2f_\beta)}+1\right)^{2}\right]^{1/2}.
\end{eqnarray}
where $\zeta=(f_\alpha-f_\beta)/f_\beta$ and $z=B(m)f_{\beta}$. The details of this demonstration are included in the appendix. Since $\zeta<0$ in the region where patterns exist, we conclude that $A(m)>1$. Indeed, this have been verified in all simulations that we have performed. We then relate this nonlinear solution to the approximated linear analysis. We can rewrite $B(m)$ as a function of  $\psi$ where $\psi$ is defined by Eq. (\ref{lins}). Then, for $\psi\rightarrow 0$ we expand $B$ as a function of $\psi$ to obtain the leading term
\begin{equation}
B(m) \propto\sqrt{\psi}\,.
\end{equation}
Here, it is noteworthy that the general solution (\ref{interaction_competition}) justifies that $\psi>0$ (as observed in the linear analysis) and we also obtain the amplitude of the waves, which is not obtained by the linear perturbation expansion.

\section{Gaussian patterns}

In the above section, we studied the wavelike patterns that arise of Eq. (\ref{interaction_competition}), for the number of crests $m\geq 3$, in which we have proposed the general analytical solution Eq. (\ref{cos}). Nevertheless, we have numerically observed the existence of modes with $m=2$ crests.

\begin{figure}[h!]
\begin{centering}
\rotatebox{-90}{\resizebox{9.0cm}{!}{\includegraphics{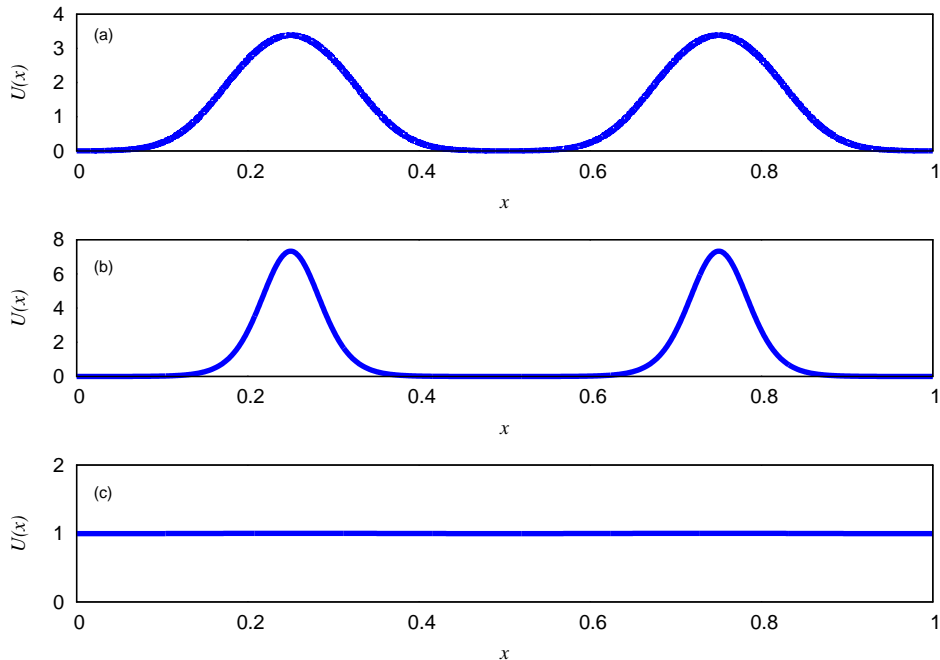}}}
\caption{Population density $U(x)$ as a function of the position $x$. Lengths are given in units of $L$, and $U(x)$ in units of $L^{-1}$, and $\alpha=0.03$ for all curves. Curve (a): $\beta=0.3$; curve (b): $\beta=0.45$; and curve (c): $\beta=0.495$.}
\label{fig.ux2}
\end{centering}
\end{figure}

In  Fig. \ref{fig.ux2} we show the steady state population density $U(x)$ as a function of the position $x$. They are for fixed $\alpha$, and they have a common characteristic, i.e., we increase $\beta$, but they do not change the number of crests. These curves  are not  described by the wavelike pattern Eq. (\ref{cos}). Indeed, curves that presents two crest are better described by combination of two identical Gaussians in the form

\begin{equation}
U(x)= C_0 +C_1[G(x-L/4)+G(x-3L/4)]\,,
\label{Gauss}
\end{equation}
where
\begin{equation}
G(x)= \frac{1}{\sqrt{2 \pi \sigma_x^2}}\exp{(-\frac{x^2 }{ 2\sigma_x^2})}.
\end{equation}

As $\beta$ increases the dispersion $\sigma_x$, first decreases and afterwards increases, $C_0$ increases monotonously towards $1$, and $C_1$ decreases monotonously towards zero until the curve becomes the uniform line $U(x)=u_0=1$. In (c) we have a very small deviation from the uniform density.
Note that for $\beta=\beta_c=1/2$ the integral in the denominator of Eq. (\ref{iterative process}) becomes just a number. In this case the integral in the denominator can not be fulfilled, unless $U(x)=u_0$, which is the uniform distribution.
The integrals for the Gaussians are more complex than in the previous case, but we  show that despite the functional form that they present, their numerical values are the same. These are the only solutions that we have found.

\begin{figure}[h!]
\rotatebox{-90}{\resizebox{7cm}{!}{\includegraphics{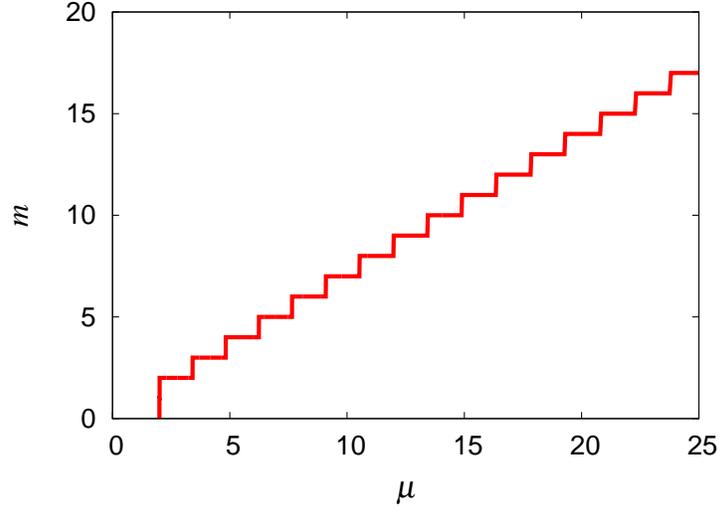}}}
\caption{ Number of crests $m$ as a function of $\mu=\beta^{-1}$.}
\label{figNc}
\end{figure}

In Fig. \ref{figNc} we plot the number of crests $m$ as a function of $\mu=\beta^{-1}$. We use $\alpha=0.01$.  Note that we start
with no pattern, and we see that $m$ increases with $\mu$ as a general characteristic. Indeed, for $0 \leq \mu \leq 2$, $m=0$, and we
have no modes, while for  $\mu^*_1  \approx 2$, $m=2$, and the modes start to appear from the background ( see Fig. 2c with $\mu=2.02$). This
corresponds to a combination of two Gaussians that holds for the region $2 < \mu < 3.430$.  As we increase $\mu$, we obtain $\mu^*_2=3.430$, where the number of crests $m$ increases from to $2$ to $3$, which corresponds to the cosine solution Eq. (\ref{cos}). Again, as $\mu$ increases, $B(m)$ increases, reaches a maximum value, decreases, becomes null, and the mode becomes unstable. Before, $B(m)$  becomes null, $B(m+1)$ starts to increases, and there is an interference between them until only the $m+1$ mode is present. This is exposed in Fig. \ref{fig.ux} for $m=5$ crests in the order (c) $\rightarrow$ (b) $\rightarrow$ (a). In general, as $\mu$ increases, $m$ increase by one unit, gradually, step by step. Note that the nonlocal integrals of Eq. (1) do not allow one single crest. This can be mathematically proved if we start with a Gaussian, there will be bifurcations to a several crests pattern, or the crest will disappear upon becoming an uniform distribution. Thus, the one crest solution is unstable.

\begin{figure}[h!]
\rotatebox{-90}{\resizebox{7cm}{!}{\includegraphics{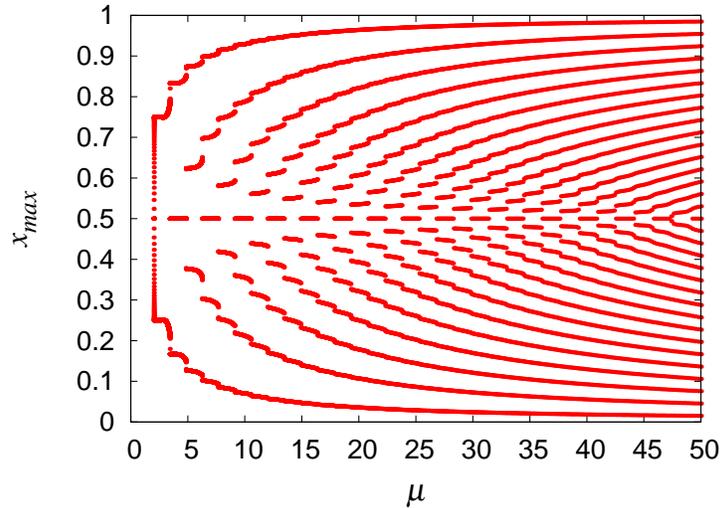}}}
\caption{ Fishbone diagram. Position of the maxima $x_{max}$ as a function of $\mu=\beta^{-1}$.}
\label{xmax}
\end{figure}

In Fig. \ref{xmax} we show the position of the maxima $x_{max}$ as function of $\mu= \beta^{-1}$. The data are the same as those of Fig. \ref{figNc}. On the left we see that as $\mu \rightarrow \mu^*_1=2$ the two crests evolve rapidly to one,  and at this point the pattern disappears. As shown in Fig. \ref{figNc}, as we increase $\mu$, we increase $m$.  This process evolves such that for a given $\mu^*_m$, $m$ changes from $m$ to $m+1$. At this point, in order to have  sufficient space to one more crest, the maxima have to adjust themselves to new positions. Note also the beautiful fractal pattern that repeats itself. For odd $m$, we have a position at the center as required by symmetry, which disappears in the next step when we go to an even number, and appears again in the next odd $m$. The whole process is like a bifurcation occurring at the center while the other maxima moves away to make space for the new one. This phenomenon is very similar to the recently reported phenomenon of peak adding in nonlinear optics \cite{Junges12}.

This process reminded us of the bifurcations of the logistic map. However, at the instability point, we do not have a periodic doubling, i.e., as shown in Fig. \ref{figNc}, we do not double $m$, we simply increase it by one.  Moreover, the limit
\begin{equation}
 \lim_{m \rightarrow \infty} \delta_m= \lim_{m \rightarrow \infty}\frac{\mu_{m+1}-\mu_{m}}{\mu_{m+2}-\mu_{m+1}},
\end{equation}
does not converge to the Feigenbaum Constant \cite{Feigenbaum},  or indeed to any other constant; rather it fluctuates.

\section{Order parameter }

Heretofore, we have defined pattern as a non-uniform process in the  population density $U(x)$.  Now, we  define it in a more quantitative way by considering the nonuniform manifestation in the density through the deviation $\Delta U(x)=U(x) - U_{NP} \neq 0$. On the other hand, $\Delta U(x)=0$ means an uniform population distribution.
The total amount of individuals in a domain $\Omega$, is given by
\begin{equation}
\label{Nt}
N(t)=\int_{\Omega} U(x,t)dx\,\,.
\end{equation}
This allows us to study the evolution of the population. Therefore, we establish that  for the long time behavior
\begin{equation}
\lim_{t\rightarrow \infty}N(t)=\begin{cases} N_{P}, & \mbox{for pattern formation }  \\  N_{NP}=1, & \mbox{otherwise.}
\end{cases}
\end{equation}
I.e., after convergence we have  $N_P$ or $N_{NP}=u_oL=1$ for pattern or no pattern formation, respectively.
Consequently, we then  define the order parameter as
\begin{equation}
\label{ro}
\rho=\int_{\Omega} \Delta U(x) dx =N-N_{NP}=\begin{cases} N_P-N_{NP} > 0, & \mbox{for pattern formation }  \\N_{NP}- N_{NP}= 0, & \mbox{otherwise}.
\end{cases}
\end{equation}
Here, $N$ is given by Eq. (\ref{Nt}).  The order parameter $\rho$ represents the total number of individuals in one phase minus the number of individuals in the homogeneous phase in such a way that it measures the extra population due to pattern formation. It is analogous to the order parameter for the liquid-vapor phase transition, i.e., $\rho=\rho_l-\rho_v$, where $\rho_l$ and $\rho_v$ stand for liquid and vapor density, respectively \cite{Huang}. Equation (\ref{ro}) can also be used for another type of population equation.

\begin{figure}[h!]
\begin{centering}
\rotatebox{-0}{\resizebox{10.0cm}{!}{\includegraphics{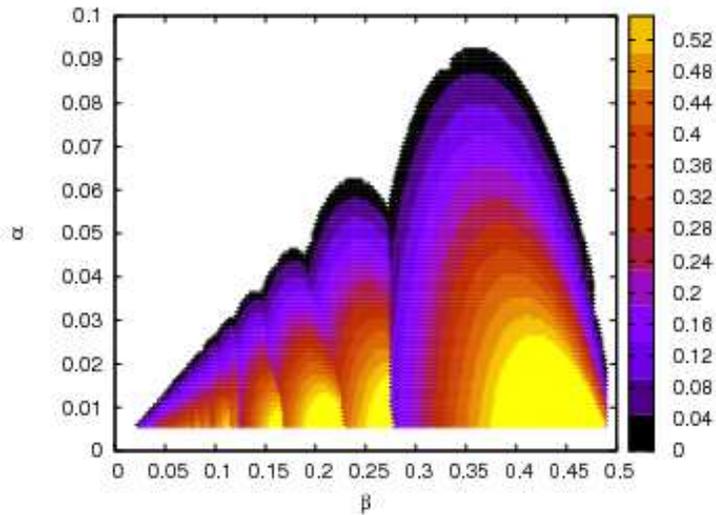}}}
\caption{The phase diagram in the $(\alpha,\beta)$ space. We show the spectra of $\rho>0$. The black line  separates the internal region where patterns exist from the external, white region, which is without patterns.}
\label{fig.map}
\end{centering}
\end{figure}

In Fig. \ref{fig.map} we plot the spectrum of $\rho(\alpha,\beta)$ in the  length parameter space $(\alpha,\beta)$.  This distinguishes the region where there is  order $\rho>0$ from that where there is no order ( $\rho=0$, white color). The   black region $0.01>\rho>0$  is the interface between the ordered (pattern), and the disordered (no pattern) phases.  A region with $\rho<0$ was not found. We shall prove later that  $\rho$ is always great than zero.

We now analyze the dependence of $\rho$ as a function of each parameter.  First, we show that by an expansion in the growth part of Eq. (\ref{interaction_competition}), we obtain a second order diffusive term with diffusion constant  $D=a \alpha^2/6$. This shows that at least for small $\alpha$,  $\alpha<<L$, one can consider $\alpha \propto T^{ 1/2}$, i.e.,  ``an absolute temperature`` of the system. In any situation, $\alpha$ can be considered to be a measure of disorder.

\begin{figure}[h!]
\begin{centering}
\rotatebox{-90}{\resizebox{7.0cm}{!}{\includegraphics{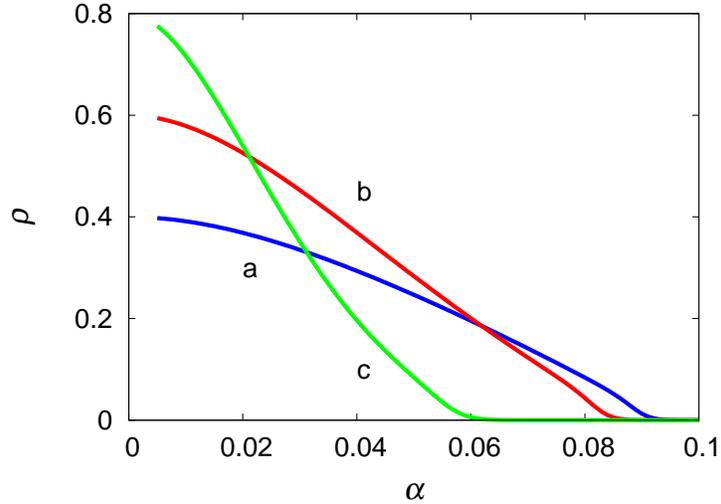}}}
\caption{The order parameter  $\rho$ as a function of the growth length  parameter $\alpha$. Here  $\rho$ is given in units of   $u_0L=1$. Curve a: $\beta=0.35$; curve b: $\beta=0.40$; and curve c: $\beta=0.45$.}
\label{fig.roal}
\end{centering}
\end{figure}

\begin{figure}[h!]
\begin{centering}
\rotatebox{-90}{\resizebox{7.0cm}{!}{\includegraphics{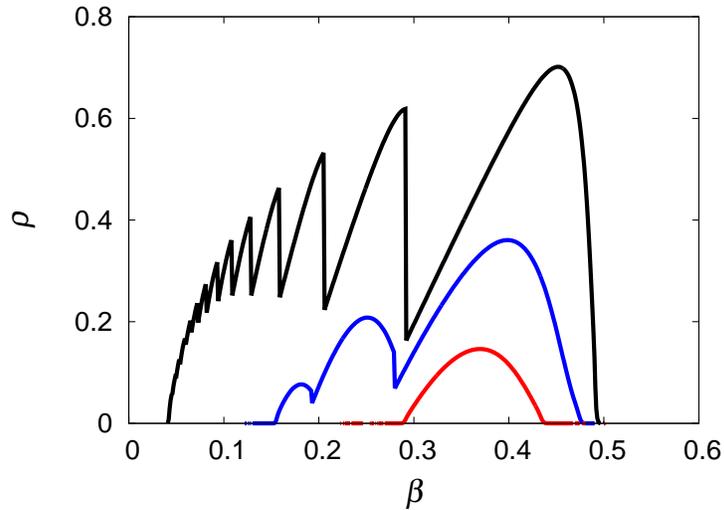}}}
\caption{The order parameter  $\rho$ as a function of the competition length  parameter $\beta$. Here $\rho(\beta)$ is given in units of   $u_0L$.  From top to bottom  $\alpha=0.01$;  $\alpha=0.04$; and $\alpha=0.07$.}
\label{fig.robeta}
\end{centering}
\end{figure}

In Fig. \ref{fig.roal}, we show the order parameter $\rho$ as a function of $\alpha$. We note that $\rho$ decreases monotonously with $\alpha$. This is very similar to the behavior of the order parameter with temperature. Consequently, the increase in $\alpha$ can be associated with a disorder in the system.  When the individuals of a niche are traveling together
they can not be far away from each other, otherwise disorder may reduce the niche. In this way we expect small $\alpha$.

In Fig. \ref{fig.robeta}, we show  the order parameter  $\rho$ as a function of the competitive length  parameter $\beta$. For the lowest curve, large $\alpha=0.7$, we see that for a specific value of $\beta$, $\rho$  starts to increase from zero, reaches a maximum, and then starts to decrease again until it becomes null. The order parameter exist only for that interval. For the upper curves, we see this behavior in a more complex way, with several curves going up and down. This unexpected behavior of $\rho$ is completely different from that exhibited in Fig. \ref{fig.roal}, and shows how rich  the dynamics is, where small changes   in the parameter space  $(\alpha,\beta)$ may yield different patterns.

 In magnetic systems the parameter space is $(T,H)$, while in the vapor-gas transition it is  $(T,P)$, where $T$ is the absolute temperature, $H$ is the external magnetic field, and $P$ is the pressure.
 Note that in our space $(\alpha,\beta)$ we associated $\alpha$ with $T$, however, we can not associate $\beta$ with $H$ or $P$ in a simple way.
  Nevertheless, we can make an analogy with the pressure in a liquid-liquid transition for complex liquids such as water, where many phases are present  \cite{Barbosa11, Barbosa15}. There a continuous increasing in pressure makes the phase disappear and to originate   a new phase. However, we believe that the behavior of  $\rho(\alpha,\beta)$   is a phenomena which needs to be better understood.

For a single mode of the wave like solution  Eq. (\ref{cos}),  we get
\begin{equation}
\label{romin}
\rho =A(m)-1=\zeta \left(1-\frac{1}{\sqrt{1-z^2}}\right)>0.
\end{equation}
Note that as we discussed previously, $\rho>0$ (see Eq. (13)). In the limit $z\rightarrow 0 $ we obtain
$\rho \propto \psi=f_\alpha-f_\beta-1 \geq 0 $,
i.e.,  the Landau criteria for stability Eq. (\ref{lins}).

 For any value of $(\alpha,\beta)$, within the cosine solutions,
the existence of order requires that $ f_{\alpha} \approx 1$ and $f_\beta <0$, and this naturally implies that $\alpha < \beta$. As $\beta \leq L/2$, we would expect to obtain $m \geq 3$ and $\rho \geq 0$, and this is precisely what we observe.  Two crests, $m=2$, are obtained by combination of two identical Gaussians (Eq. (\ref{Gauss})).  Again for this case  $\rho \geq 0 $.

 Now, we analyze together  Figs. \ref{figNc} and \ref{fig.robeta}.  In Fig. \ref{figNc},  we show the number of crests $m$ as a function  of $\mu= \beta^{-1}$. Here, we use $\alpha=0.01$, which corresponds to the upper curve of Fig. \ref{fig.robeta}.
 We start with two modes, for $\mu  \geq 2$ i.e., $\beta \le 0.5$, and this corresponds to a combination of two Gaussians.   As $\mu$ increases we reach $m=3$, which now corresponds to the cosine solution Eq. (\ref{cos}).   Again, as $\mu$ increases $B_m$ increases,  reaches a maximum value, and then decreases until it becomes null and the mode unstable. Before  $B_m$  becomes null   $B_{m+1}$ starts to increases,  and there is an interference between  the modes. Finally,the $m+1$ mode dominates, and the process continues.
 This is exposed in Fig. \ref{fig.ux} for $m=5$ in the order (c) $\rightarrow$ (b)  $\rightarrow$ (a). In other words, it shows the transition from $m=5$ to $m=6$. In general as $\mu$ increases $m$ increase by one , step by step.  These phenomena are very similar to recently reported phenomena in nonlinear optics where self-pulsations were observed in a CO2 laser with feedback display two types of recurrent period discontinuities when the control parameters were changed.
Periodic self-pulsations emerge organized in wide adjacent phases wherein oscillations differ by a constant number of peaks in their period \cite{Junges12}.
It is quite interesting to see that in our case the number of crests does not exhibit any lacuna.  This, consequently strengthens our argument  that we have seen all modes and that there are only cosines and Gaussian solutions.

 In the region $\alpha \geq 0.07$,  $m = 0$ or $2$, independently of $\beta$.  For  $\alpha = 0.07$ (the lowest curve of Fig. \ref{fig.robeta}, is a simple half loop, i.e., our unity structure. Every half loop of Fig. \ref{fig.robeta} corresponds to a fixed number of crests, and the full curve is a superposition of these half loops.
Now, the reason for the up and down of $\rho$ in Fig. \ref{fig.robeta} become clear. In the lowest one, the half loop, for that value of $\alpha$
as we increase $\beta$ we start to have a pattern with two crests, i.e., $m=2$. The order parameter  increases as the pattern arises
until it reaches a maximum, and then it decreases to a null value when it is not possible to have the existence of a new pattern with one crest. For the upper curves, every half loop corresponds to a given $m$ that decreases as $\beta$ increases. In general,
in the interval, $\beta^*_{m+1} < \beta < \beta^*_m $, or equivalently $\mu*_m < \mu* < \mu*_{m+1}$,
$\rho$ first increases to a maximum and then decreases. However, if $m \geq 2$ there will be another pattern and another half loop will start. Consequently, at this transition the order parameter  does not drop to zero. Indeed, the new loop does not start from a null value, rather; it  superposes the previous one. At the moment we have reached $m=2$, as in the lowest curve, the transition goes directly to the uniform solution, i.e., to $U(x)=u_0=a/b$, and $\rho=0$.

\section{entropy }
Finally, in this section we try to extend our thermodynamic analogy.  In order to do that we obtain the entropy, since it is one of the most important thermodynamic  functions, and it would be useful for our analysis. We expect that as we have a more ordered system, as occurs when patterns arises, we should observe a decrease of entropy. The definition of  entropy,  allows us to describe the thermodynamics of  unusual systems. For example, the synchronization of the Kuramoto model with multiplicative noise using a thermodynamic formulation was recently discussed \cite{Pinto16}. This is a simplification of the usual dynamic point of view. Our major interest here is to relate  entropy and  the order parameter. We begin by defining the Gibbs entropy in units of the Boltzmann constant $k_B$ as
\begin{equation}
 S = - \sum_j P_j \ln P_j,
\end{equation}
where $P_j= U(x_j)\Delta x/N$ is the probability of finding the individuals in the region $x_j- \Delta x/2 < x < x_j+ \Delta x/2$.
For an uniform distribution  $S_{NP}= \ln(\frac{L}{\Delta x})=\ln(N)$.  As in any situation in classical systems it depends on  the division size $\Delta x=L/N$. Thus, for an inhomogeneous distribution the entropy  is
\begin{equation}
\label{entropy}
 S(\rho)= S_{NP} + \int_{\Omega} \phi(\rho,x)\ln[1-\phi(\rho,x)]dx\,\,,
\end{equation}
where we obtain $\phi$ as a function of the order parameter as
\begin{equation}
\label{phi}
\phi(\rho,x)=\frac{\rho-L\Delta U(x)}{\rho+LU_{NP}}.
\end{equation}
Since the minimum value of $\Delta U(x)$ is $-u_0$, the maximum value of $ \phi(\rho,x)$ is $1$, and consequently the integral is well defined and negative. For every $\rho$ we have a $\Delta S=S_{NP}-S(\rho)>0$, which corresponds to minus the integral in Eq. (\ref{entropy}) and is the amount of entropy
that the system releases
to form a more organized state. This quantity is independent of the division $\Delta x$.

\begin{figure}[h!]
\begin{centering}
\rotatebox{-90}{\resizebox{7.0cm}{!}{\includegraphics{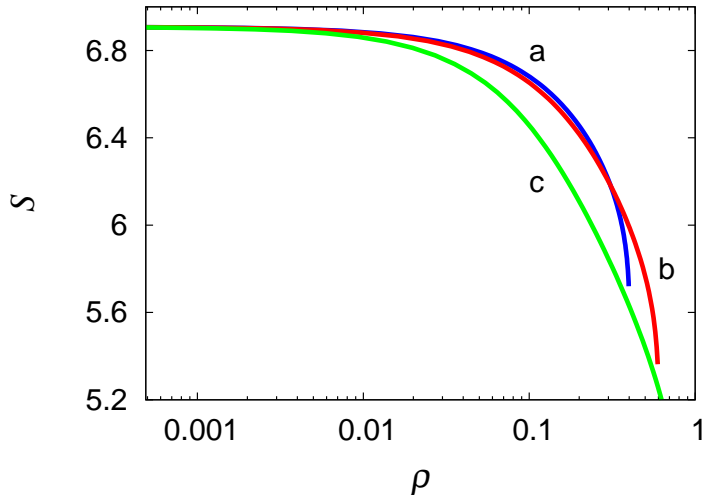}}}
\caption{Entropy $S$ as a function of the order parameter  $\rho$. Here we fix $\beta$ and we change $\alpha$. Data  as in Fig. \ref{fig.roal}.}
\label{fig.sal}
\end{centering}
\end{figure}

In Fig. \ref{fig.sal}, we show  the entropy $S(\rho)$ as a function of the order parameter  $\rho$ for a fixed $\beta$.
The data are as shown in Fig. \ref{fig.roal}. As expected, we see the decreasing of entropy with an increasing in the  order parameter. A plot of the entropy as a function of $\alpha$ for fixed $\beta$ yields a similar pattern where $S(\alpha)$ decreases monotonically with $\alpha$. Again, considering $\alpha$ as a disorder parameter similar to temperature, this is as expected.

\begin{figure}
\begin{centering}
\rotatebox{-90}{\resizebox{7.0cm}{!}{\includegraphics{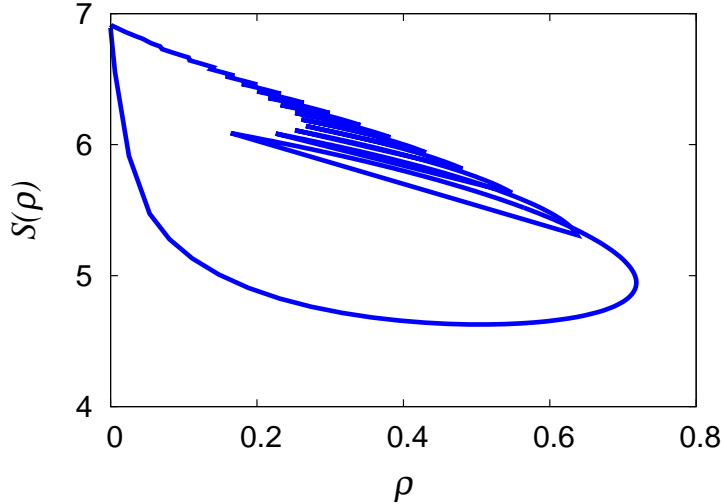}}}
\caption{ Entropy $S(\rho)$ as a function of the order parameter  $\rho$ for fixed $\alpha$.}
\label{fig.Sro}
\end{centering}
\end{figure}

In Fig. \ref{fig.Sro}, we show  the entropy $S(\rho)$ as a function of the order parameter  $\rho$ for fixed $\alpha$. The data are that of the upper curve of Fig. \ref{fig.robeta}.
 We do not see a monotonous decreasing of entropy as the  order parameter increases, as in  Fig. \ref{fig.sal}; rather, we see  a behavior that is more rich and complex, similar to  that  of  Fig.  \ref{fig.robeta}.
Note that every half loop of  Fig. \ref{fig.robeta} appears here as a loop where the entropy increases and decreases. For every $m$, the maximum of $\rho$ corresponds to a minimum of the entropy. For large $m$, the passage $m+1 \rightarrow m$ occurs with very little change in entropy; as $m$ becomes small the change demands a  large variation in the entropy. This change in entropy as we change $\beta$, for fixed $\alpha$, i.e., a fixed ''temperature``, is the latent change in entropy for a phase transition. The transition here is not only the order-disordered transition, but also, the order-order transition when we move from a pattern of $m+1$ crests to a pattern of  $ m \leq 2$ crests.

One major question is ``Are these values for the order parameter, Fig. \ref{fig.robeta}, and   entropy Fig. \ref{fig.Sro}, one possible reality or spurious results from the model?''  If one thinks about an ecological model out of equilibrium, where growth, competition, diffusion, dispersion, and nonlocality are present, it is quite reasonable to obtain multiple possible values for entropy. One value is obtained for each $m$, i.e., for each given possibility of a stationary mode.  In  dynamical evolution, the possibility of transition from one stationary structure to another is easier if many possibilities exist.
Part of the  diversity and complexity we see in nature is exhibited in this simple model.

We now turn to the central problem in the analysis of pattern formation.
Why should a  biological system make a pattern? Our model allows us to obtain answers from a simple perspective.
  It is well established  that nature favors diversity. In addiction it would be expected that patterns would reduce the entropy of the system. This is exactly what we find here. Moreover, in Eq. (\ref{romin}), as well in  Figs. \ref{fig.map}, \ref{fig.roal}, and \ref{fig.robeta}, it is clear  that $\rho >0$. Thus, the existence of patterns means that a large number of individuals can live in a given  ecological niche,
i.e.,  a  number larger than that provided by  a uniform density.
Thus, it would be expected that if we have an evolutionary  population dynamics, where the systems can choose through natural selection of  $(\alpha,\beta)$ they will evolve to choose values for these parameters such that $\rho = N_P - N_{NP} >0$ is large, where    $ N_P > N_{NP} $. Thus, as shown by this minimal model, this could be a good reason for nature to make patterns.
Furthermore, as $U(x)>0$, then $\phi(\rho,x)<1$, and therefore the integral in Eq. (\ref{entropy}) is always negative. Thus $S(\rho)$ must be a decreasing function of $\rho$. We also note that $\phi(0,x)=0$ implies $S(0)=S_{NP}$.  Consequently, as the entropy is a decreasing function of $\rho$, and since any pattern will have $S_{P}<S_{NP}$, we conclude that patterns will necessarily satisfy $\rho>0$. Therefore, this confirms mathematically the rule we have previously established that for nature to have more individuals in a given domain it has to create patterns.

\section{Conclusions and outlook}

In conclusion, we have developed a nonlinear integro-differential equation with double nonlocal terms for the population density function. In this formulation, we have two kernels and two correlation lengths $\alpha$ for growth,  and another one, $\beta$, for competition. We have numerically solved this equation studied in detail how the population density evolves to a state of inhomogeneous density (pattern) or to a uniform density (no pattern). We also obtained analytical  solutions for  the steady state as a cosine function for three peaks or more, and for two peaks as a double Gaussian. The cosine type  solutions are surprising for this complex nonlocal nonlinear integro-differential equation.

We study the stability of the modes  for a given density $U(x)$, and we have described how a transition from $m$ to a $m+1$ crests occurs as the competition parameter decreases. The breaking of  the mode stability does not lead us to a bifurcation, rather to the addition of a new crest. This is similar to a recently observed  phenomenon   in nonlinear optics \cite{Junges12} known as a pick adding. We define an order parameter $\rho$  that exhibits some very rich phenomena.
The evolution of $\rho$ with $\alpha$ is similar to the evolution of the order parameter in the liquid-vapor transition with temperature. The evolution of $\rho$ with $\beta$ shows that not only do the transitions between order and disorder occur, but also that the transition from a pattern of $m$ crest to a pattern of $m \pm 1$ crest may occur, if  the latent entropy conditions are appropriated.

Finally, we believe that these results strengthen the relationship between patterns formation and phase transitions in complex systems with many phases, such as water \cite{Barbosa11,Barbosa15}, as we have mentioned before . The possibility of analysis of systems in higher dimensions, e.g., $d=2,3$, and, non-stationary states \cite{Hallatschek09, Liua16}, and, the inclusion of  noise \cite{Longa96,Bocaletti02,Ciesla01,Zhou02,Morgado07}, and competition between several species  \cite{Valenti04,Hatton15}  show that we are just at the beginning, and  further investigations in this area will likely lead to further important breakthroughs.

\section{Acknowledgments}

 We acknowledge the support of the Conselho Nacional de Desenvolvimento Cient\'ifico e Tecnol\'ogico(CNPq), Coordena\c{c}\~ao de Aperfei\c{c}oamento de Pessoal Superior (CAPES), and a Funda\c{c}\~ao de Apoio a Pesquisa do Distrito Federal (FAPDF) Brazil. Two of us (RMSF and FAO) would like to thank Professor Hyunggyu Park for his hospitality during their visit to the Korean Institute for Advanced  Study (KIAS).

\appendix
\section{Determination of the parameters A and B}

Now, we  call $U_n(x)$ the solutions (\ref{cos}). We then  solve iteratively Eq.  (\ref{iterative process}) to obtain
\begin{equation}
\label{UU}
U_{n+1}(x)=1+\zeta-\frac{\zeta}{1+z\cos(kx)}\,,
\end{equation}
where $n$ is the order of the iteration, and we have defined
\begin{equation}
\zeta=\frac{f_\alpha-f_\beta}{f_\beta},\quad z=Bf_{\beta},\quad f_{\sigma}=\frac{\sin(k_m \sigma)}{k_m \sigma}\,.
\end{equation}
We expand (\ref{UU}) in a series of $z\cos(kx)$, and we replace the higher order terms by their average in order to be coherent with Eq. (\ref{cos}), then
\begin{equation}
U_{n+1}(x)=1+\zeta-\zeta\lambda +\zeta z\lambda\cos(kx)\,,
\label{eq.6}
\end{equation}
where
\begin{equation}
\lambda=\sum_{n}\lambda_{n}=\frac{\tilde{\lambda}}{\sqrt{1-\phi}}\,,
\end{equation}
\begin{equation}
\lambda_{n}=\frac{z^{2n}}{L}\int_{0}^{L}\cos(kx)^{2n}dx\,,
\label{lamb}
\end{equation}
$\tilde{\lambda}=2/(2-z^2)$ and $\phi=(z^2\tilde{\lambda})^2/4$. To find the parameter $A(M)$, we use (\ref{cos}), which gives
\begin{equation}
\label{AA}
A(m)=1+\zeta \left(1-\frac{1}{\sqrt{1-z^2}}\right)\,.
\end{equation}

To obtain the parameter $B(m)$ we start from the discrete form of the function $\eta$ in Eq. (\ref{stability})
\begin{eqnarray}
\eta&=&\int^l_0[U_n(x)-U_{(n+1)}(x)]^2dx\nonumber\\
\nonumber\\
&=&A(m)^2\left(1+\frac{B^2}{2}\right)
-2A(m)\left[1-\zeta\left(\frac{1}{f_\beta}-1\right)\left(1-\frac{1}{\sqrt{1-z^2}}\right)\right],
\end{eqnarray}
 In order to minimize the parameter $\eta$ in relation to $A$, we take
\begin{equation}
\frac{\partial\eta}{\partial A(m)}=2A(m)\left(1+\frac{B^2}{2}\right)-2\left[1-\zeta\left(\frac{1}{f_\beta}-1\right)\left(1-\frac{1}{\sqrt{1-z^2}}\right)\right]=0,
\end{equation}
then
\begin{equation}
A(m)= \frac{1}{{1+{B^2}/{2}}}\left[1-\zeta\left(\frac{1}{f_\beta}-1\right)\left(1-\frac{1}{\sqrt{1-z^2}}\right)\right].
\label{eq.a2}
\end{equation}
Equating the expression (\ref{eq.a2}) to the expression (\ref{AA}) and defining $\Phi=(1-z^2)^{1/2}$, we obtain
\begin{equation}
\Phi^3(1+\zeta)-\Phi^2\zeta-\Phi(1+2\zeta f_\beta+\zeta)+\zeta(1+2f_\beta)=0.
\end{equation}
Note that one of the solutions is $\Phi=1$ or $z=B(m)f_{\beta}=0$, which implies $B(m)=0$ and $A(m)=1$. Indeed, this represents the no pattern solution, as expected. Two other solutions are
\begin{equation}
\label{PP}
\Phi_{\pm}=-\frac{1}{2(1+\zeta)}\pm\frac{\sqrt{1+4\zeta(\zeta+2f_\beta+2f_\beta\zeta)}}{2(1+\zeta)}.
\end{equation}
Now we can solve Eq. (\ref{PP}) for $B$, which results in
\begin{equation}
B(m)=\pm\frac{1}{2(1+\zeta)f_\beta}\left[4(1+\zeta)^2-\left(\sqrt{1+4\zeta(1+\zeta)(1+2f_\beta)}+1\right)^{2}\right]^{1/2}.
\label{eq.b}
\end{equation}
The numerical solution converges to a positive or negative $B(m)$, and we have no control for it. This completes the demonstration of the parameters $A(m)$ and $B(m)$ for Eq. (\ref{cos}).

\end{document}